\input harvmac





\def\underarrow#1{\vbox{\ialign{##\crcr$\hfil\displaystyle
 
{#1}\hfil$\crcr\noalign{\kern1pt\nointerlineskip}$\longrightarrow$\crcr}}}
%
\def\tilde{\widetilde}
\def\bar{\overline}

%

\font\cmss=cmss10
\font\cmsss=cmss10 at 7pt
\def\rlx{\relax\leavevmode}
\def\inbar{\vrule height1.5ex width.4pt depth0pt}
\def\IC{\relax\,\hbox{$\inbar\kern-.3em{\rm C}$}}
\def\IN{\relax{\rm I\kern-.18em N}}
\def\IP{\relax{\rm I\kern-.18em P}}
\def\IR{\relax{\rm I\kern-.18em R}}
\def\IC{{\relax\hbox{$\inbar\kern-.3em{\rm C}$}}}
\def\IZ{\relax\ifmmode\mathchoice
{\hbox{\cmss Z\kern-.4em Z}}{\hbox{\cmss Z\kern-.4em Z}}
{\lower.9pt\hbox{\cmsss Z\kern-.4em Z}}
{\lower1.2pt\hbox{\cmsss Z\kern-.4em Z}}\else{\cmss Z\kern-.4em
Z}\fi}
\def\IH{\relax{\rm I\kern-.18em H}}
\def\ZZ{\rlx\leavevmode\ifmmode\mathchoice{\hbox{\cmss Z\kern-.4em Z}}
 {\hbox{\cmss Z\kern-.4em Z}}{\lower.9pt\hbox{\cmsss Z\kern-.36em Z}}
 {\lower1.2pt\hbox{\cmsss Z\kern-.36em Z}}\else{\cmss Z\kern-.4em
 Z}\fi}
\def\narrowplus{\kern -.04truein + \kern -.03truein}
\def\narrowminus{- \kern -.04truein}
\def\narrowminussub{\kern -.02truein - \kern -.01truein}

\def\frac#1#2{{#1\over #2}}

\def\CD{{\cal D}}

\def\IZ{\relax\ifmmode\mathchoice
{\hbox{\cmss Z\kern-.4em Z}}{\hbox{\cmss Z\kern-.4em Z}}
{\lower.9pt\hbox{\cmsss Z\kern-.4em Z}}
{\lower1.2pt\hbox{\cmsss Z\kern-.4em Z}}\else{\cmss Z\kern-.4em
Z}\fi}

%
%
\def\eqnn#1{\xdef #1{(\secsym\the\meqno)}\writedef{#1\leftbracket#1}%
\global\advance\meqno by1\wrlabeL#1}
\def\eqna#1{\xdef #1##1{\hbox{$(\secsym\the\meqno##1)$}}
\writedef{#1\numbersign1\leftbracket#1{\numbersign1}}%
\global\advance\meqno by1\wrlabeL{#1$\{\}$}}
\def\eqn#1#2{\xdef #1{(\secsym\the\meqno)}\writedef{#1\leftbracket#1}%
\global\advance\meqno by1$$#2\eqno#1\eqlabeL#1$$}


\nref\bfss{T. Banks, W. Fischler, S. Shenker and L. Susskind,
``M-Theory as a Matrix Model: A Conjecture'',
hep-th/9610043,
Phys. Rev. D55, 5112, 1997} 

\nref\juan{J. Maldacena, ``The Large N Limit of Superconformal 
Field Theories and Supergravity'', hep-th/9711200, Adv. Theor. Math. 
Phys.2,231,1998}


\nref\twozera{O. Aharony, M. Berkooz, S. Kachru, 
N. Seiberg and E. Silverstein,
``Matrix Description of Interacting Theories in Six Dimensions'',
hep-th/9707079,
Adv.Theor.Math.Phys.1, 148, 1998}

\nref\twozerb{O. Aharony, M. Berkooz and N. Seiberg,
``Light Cone Description of (2,0) Superconformal Theories in 
six dimensions'', hep-th/9712117, Adv. Theor. math. Phys.2, 119, 1998}

\nref\nak{H. Nakajima, ``Resolution of Moduli Spaces of Ideal Instantons
on $R^4$'', in ``Topology, Geometry and Field Theory'', 
ed. Fukaya, Furuta, Kohno and Kotschick, World Scientific}

\nref\edltls{E. Witten, 
``On the Conformal Field Theory of the Higgs Branch'',
hep-th/9707093,
JHEP 9707, 003, 1997}

\nref\md{M. Berkooz and M. Douglas,
``Five-Branes in M(atrix) Theory'',
hep-th/9610236, Phys. Lett. B395, 196, 1997}

\nref\raem{R. Entin and E. Diaconescu,
``A Nonrenormalization Theorem for the D=1, N=8 Vector Multiplet'',
hep-th/9706059, Phys. Rev. D56, 8045, 1997}

\nref\natidec{N. Seiberg,
``New Theories in Six Dimension and Matrix Description of M-Theory on 
$T^5$ and $T^5/Z_2$'',
hep-th/9705221, Phys. Lett. B408, 98, 1997}

\nref\dvv{R. Dijkgraaf, E. Verlinde and H. Verlinde,
``BPS Spectrum of Fivebranes and Black Hole Entropy'', hep-th/9603126,
Nucl. Phys. B 486, 77, 1997; ``BPS Quantization of the Fivebrane'',
hep-th/9604055, Nucl. Phys. B 486, 89, 1997;
``5D Black Holes and Matrix Strings'', hep-th/9704018,
Nucl. Phys. B 506, 121, 1997}

\nref\brs{M. Berkooz, M. Rozali and N. Seiberg,
``Matrix Description of M-Theory on $T^4$ and $T^5$''
hep-th/9704089, Phys. Lett.B 408, 105, 1997}

\nref\ltlnh{O. Aharony, M. Berkooz, D. Kutasov and N. Seiberg, 
``Linear Dilatons, NS 5-branes and Holography'', hep-th/9808149,
JHEP 9810:004,1998
} 

\nref\chs{C. Callan, J. Harvey and A. Strominger, 
``Supersymmetric String Solitons'', hep-th/9112030, in Trieste 1991,
Proceedings, String Theory and Quantum Gravity, 1991, 208}

\nref\lngstrng{L. Motl, ``Proposals on Non-Perturbative Superstring
Interactions'', hep-th/9701025;
T. Banks and N, Seiberg, ``String from Matrices'',
hep-th/9702187, Nucl. Phys. B 497, 41, 1997;
R. Dijkgraaf, E. Verlinde and H. Verlinde,
``Matrix String Theory'',
hep-th/9703030, Nucl. Phys. B 500, 43, 1997}

\nref\insta{N. Dorey, T.J. Hollowood, V.V. Khoze, M.P. Mattis and
S. Vandoren, ``Multi-Instanton Calculus and the AdS/CFT Correspondence
in ${\cal N}=4$ Superconformal Field Theories'', hep-th/9901128}

\nref\instb{N. Dorey, V.V. Khoze and M.P. Mattis,
``Multi-Instanton Calculus and ${cal N}=2$ Supersymmetric Gauge Theory'',
hep-th/9603136;
V.V. Khoze, M.P. Mattis and M.J. Slater, 
``The Instanton Hunter's Guide to supersymmetric U(N) Gauge Theories'',
hep-th/9404009}

\nref\uvir{L. Susskind and E. Witten, 
``The Holographic Bound in Anti-De Sitter Space'',
hep-th/9805114}

\nref\bndstt{S. Sethi and M. Stern,
``D-Brane Bound State Redux'', hep-th/9705046, Comm. Math. Phys. 194,
675,1998}

\nref\ansanj{A. Jevicki and S. Ramgoolam,
``Non-commutative gravity from the AdS/CFT Correspondence'',
hep-th/9902059}

\nref\ikkt{N. Ishibashi, H. Kawai, Y. Kitazawa, A. Tsuchiya,
``A Large N Reduced Model as Superstring'', hep-th/9612115,
Nucl. Phys. B498,467,1998}

\nref\orisav{O.J. Ganor and S. Sethi, ``New Perspectives on Yang-Mills 
Theories with Sixteen Supercharges'', hep-th/8712071, JHEP 9801:007,1998;
A. Kapustin and S. Sethi,
``The Higgs Branch of Impurity Theories'',
hep-th/9804027,
Adv. Theor. Math. Phys. 2, 571, 1998}

\nref\forthcoming{In preparation}

\nref\natied{N. Seiberg and E. Witten,
``The D1/D5 System and Singular CFT'',
hep-th/9903224, JHEP 9904, 017, 1999
}


\nref\ncg{A. Connes, M.R. Douglas and A. Schwarz,
``Noncommutative Geometry and Matrix Theory: Compactification on Tori'',
hep-th/9711162, JHEP 9802:003, 1998;
M.R. Douglas and C. Hull,
``D-Branes and the Non-commutative Torus'',
hep-th/9711165, JHEP 9802:009, 1998}

\nref\mp{D. Marolf and A.W. Peet, ``Brane Baldness vs. Superselection 
Sectors'', hep-th/9903213, Phys. Rev. D60, 1999}

\nref\ah{H. Awata and S. Hirano, ``AdS(7)/CFT(6) Correspondence and Matrix 
Models of M5-Branes'', hep-th/9812218, Adv. Theor. Math. Phys.3,147,1999}


\def\al{{\alpha}}
\def\be{{\beta}}
\def\ga{{\gamma}}

\def\ga{{\gamma}}
\def\be{{\beta}}
\def\al{{\alpha}}
\def\et#1#2{{\eta^{#1}_{#2}}}

\def\eps{{\epsilon}}


\Title{\vbox{\hbox{hep-th/9907100}\hbox{IASSNS-HEP-99/67}\hbox{PUPT-1879}}
} {\vbox{\centerline{Matrix Theory, AdS/CFT and Higgs-Coulomb}
\centerline{}
\centerline{Equivalence}}}
\smallskip
\centerline{Micha Berkooz\footnote{$^1$} {berkooz@sns.ias.edu} and
Herman Verlinde\footnote{$^2$} {verlinde@feynman.princeton.edu} }
\vskip 0.09in
\medskip\centerline{$^1$\it School of Natural Sciences, Institute
for Advanced Study, Princeton, NJ 08540, USA}
\medskip\centerline{$^2$ \it Joseph Henry Laboratories, 
Princeton University,
Princeton, NJ 08544, USA}
\medskip\centerline{$^2$ \it  Institute for Theoretical Physics,
University of Amsterdam}
\centerline{\it Valckenierstraat 65, 1018 XE Amsterdam, Netherlands}


\vskip 0.30in

\centerline{\bf Abstract}

We discuss the relation between the Matrix theory definitions of a
class of decoupled theories and their AdS/CFT description in terms of
the corresponding near-horizon geometry. The near horizon geometry,
naively part of the Coulomb branch, is embedded in the Higgs branch
via a natural change of variables. The principles of the map apply to
all DLCQ descriptions in terms of hyper-K\"ahler quotients, such as
the ADHM quantum mechanics for the D1-D5 system. We then focus the
(2,0) field theory, and obtain an explicit mapping from all states in
the $N_0=1$ momentum sector of $N_4$ M5-branes to states in (a DLCQ
version of) $AdS_7\times S^4$. We show that, even for a single
D0-brane, the space-time coordinates become non-commuting variables,
suggesting an inherent non-commutativity of space-time in the presence
of field strengths even for theories with gravity.

\vskip 0.05in

\Date{July 1999}
\vfill\eject

\def\bnu{{{\bar\nu}}}
\def\wvfnc#1{{ {\vert{#1}>} }}


\newsec{Introduction}

By now we have several non-perturbative formulations of M-theory and
String theory in different backgrounds or kinematical set-ups,
primarily in the frame work of Matrix theory \bfss\ and the AdS/CFT
correspondence \juan. However it is disappointing, albeit in a merry
way, to go from a situation without any non-perturbative formulations
to a situation with many background-dependent such formulations. One
would therefore like a sequence of maps from one formulation to
another.  This paper is a preliminary study into the relation between
the AdS/CFT correspondence and Matrix Theory in which we will suggest
a derivation of the former given the latter.

More concretely, both Matrix theory and the AdS/CFT correspondence
(and its generalizations) allow us to write down a description of
theories which can be obtained from solitons in string theory/M-theory
by decoupling gravity. The range of usefulness of the two methods is
different. The advantages of the AdS/CFT duality is that it can be
exhibited for a large class of field theories/gravity backgrounds, the
formulation exhibits all the symmetries of the theory manifestly, and
one can derive some qualitative properties of field theories from
it. Its drawback is that obtaining exact quantitative results is
typically possible only in extreme regimes of the parameters space of
the field theory such that the gravity background is manageable, which
is infrequent. The Matrix theory of decoupled theories on solitons has
its limitations as well. One is restricted to a small class of theories,
one looses some of the symmetries and one still needs to take the
large null-momentum limit, which often makes the problem complicated. But
when it exists, it exists just as well for cases where the
supergravity is not weakly coupled.  For example, using the DLCQ of
the (2,0) field theory it is easy to count precisely
\twozerb\ chiral operators for every number of 5-branes but not to
compute their OPE. Whereas using the AdS/CFT description it is easy to
compute the 3-point function of these operators in the extreme case of
a very large number of 5-branes, but not to count them precisely or
compute the OPE for a finite number of 5-branes. A map between the two
descriptions will perhaps enable us to enjoy the advantages of both
systems, might allow us to borrow new tools from one description to be
used in the other, or might teach us altogether new things.

In this paper we will suggest that it is natural to consider the
AdS/CFT correspondence within the framework of Matrix theory. We
emphasize that this direction is opposite to what was previously done
in the literature, which is to interpret Matrix theory as a special
case of an AdS/CFT-type correspondence. What we will show is that one
can clearly identify the near horizon description within the DLCQ of
the decoupled field theory. Thus, if one accepts these DLCQ
conjectures, then the AdS/CFT correspondence is essentially a change
of basis in the DLCQ Hilbert space. In this paper we will discuss some
basic aspects of this identification and change of basis, and leave
its more extensive elaboration to the future.

In terms of the DLCQ of the decoupled theories on the brane, the
problem can be phrased in terms of the relations between the Coulomb
and Higgs branches of certain (other) field theories. These field
theories are those provided by Matrix theory as means of describing
the dynamics on the brane prior to the decoupling from the bulk gravity;
we will refer to them as the full DLCQ theories. These
theories typically have a Coulomb branch, which describes gravity away
from the brane, and a Higgs branch, which describes excitations inside
the brane. The decoupling limit is, in the language of
the full DLCQ theory, the limit in which the Higgs branch and the
Coulomb branch decouple \edltls\twozerb. 

The puzzle now is the following. The theory on the brane has a
description in terms of gravity in the near horizon region of the
soliton. Semi-classically, however, this region seems to be part of
the Coulomb branch of the full DLCQ theory since the distance of the
excitation from the brane is not strictly zero. Admittedly, it is the
tip of the Coulomb branch where it touches the Higgs branch, but
nevertheless it seems we never enter the Higgs branch.  In order to
reconcile these points of view, i.e., to find the near-horizon
geometry in the Higgs branch, one needs to find some kind of
equivalence between the tip of the Coulomb branch and the entire Higgs
branch in these systems. Of course the problem is in the semiclassical
statement and one can excuse oneself by appealing to large quantum
fluctuations, but still one would like to do better. This is the
purpose of this paper.

Once we have identified the dynamics of gravitons in the near horizon
geometry in terms of dynamics on the Higgs branch, we have, in effect,
derived the AdS/CFT duality from Matrix theory.

We will present in this paper only the tip of the iceberg on this map
and its properties. In section two we present the general idea,
although we will cast it a form suitable for the D0-D4 system and the
D1-D5 system, which give rise to the DLCQ of the (2,0) field theory
\twozera\ and the DLCQ of the little string theory
\twozera\edltls\ respectively. Section 3 is a technical note on the map. 
In section 4 we begin focusing on the DLCQ of the (2,0) field theory
for a single unit of momentum along the null direction. Although in
Matrix theory one is instructed to take a large number of units of
such momenta, already at the single unit level we will start seeing
interesting effects. Section 4 sets up the description of this model,
and of what we will call ``the reduced model'' which is a flavor
invariant version of the ADHM quantum mechanics and which will be
intimately related to the near horizon geometry. Section 5 discusses
the other side of the equation, i.e., the
DLCQ of $AdS_7\times S_4$ (which is the supergravity dual of the
$(2,0)$ field theory). Again we specialize to the case of a single unit
of momentum. Section 6 finally discusses the details of the map and
some of its features. In particular we will see the significant
appearance of a non-commutative\foot{We will use this term somewhat
loosely. What we mean precisely will be made clear below.} version of
$S^4$.


\newsec{The Near Horizon Limit in Matrix Theory}

\subsec{Decoupling in low-dimensional SYM}

In \twozera\edltls\ it was argued that the DLCQ of the (2,0) field
theory and of the ``little string theory'' are a 1+0 and a 1+1
(respectively) sigma models on certain Higgs branches (which describe
instanton moduli spaces). Since both cases require a longitudinal
5-brane \md, the Higgs branch in question is in both cases the
dimensional reduction of 4D ${\cal N}=2$ theories with $U(N_0)$ gauge
group, a hypermultiplet in the adjoint and hypermultiplets in a
bi-fundamental of the gauge group and an $U(N_4)$ flavor symmetry. We
will discuss below each case at greater detail, however the basic
scaling that generates the Higgs/Coulomb map is common to both and
will be the topic of this section.

In our discussion here we will be schematic (we will not write
all the couplings and quantum numbers) \foot{More
precise statements can be found below, or in the above
references.}. Prior to decoupling, schematically, the fields in the
Lagrangian are:

\eqn\flds{\matrix{
Y_1&Vector\ multiplets\ scalars\cr
\theta'&\ Vector\ multiplets\ fermions\cr
H&Adjoint\ hypermultiplet\ boson\cr
\theta&\ Adjoint\ hypermultiplet\ fermions\cr
  Q&Bi-fundamental\ hypermultiplets\ bosons\cr
\mu&Bi-fundamental\ hypermultiplets\ fermions
}} as well as gauge fields (in the same multiplet as $Y_1$ and
$\theta'$), and the SYM action is
\eqn\symd{  \int d^d\sigma
{1\over g_{ym}^2}
\bigl( (\CD Y_1)^2+[Y_1,Y_1]^2 + 
\theta(\CD \theta+[Y_1,\theta]) \bigr)+}
$$ + (\CD H)^2 + [Y_1,H]^2 + \theta'(\CD \theta'+[Y_1,\theta'])+$$
$$ + (\CD Q)^2 + (Y_1Q )^2 + \mu(\CD+Y_1)\mu + g_{ym}^2
([H,H]+Q^2)^2$$
where $d=1,2$.  The dimensionality of the hypermultiplet fields
changes with dimension, but the Coulomb branch coordinate $Y_1$ is
always of dimension 1 (hence the notation $Y_1$).

In the analysis in \twozera\edltls\ one takes $g_{ym}^2\rightarrow
\infty$ keeping the energy fixed (or equivalently flows to the IR 
for fixed $g^2_{ym}$). If we are on the Higgs branch, then the mass of
the Higgsed gauge bosons goes to infinity and they decouple, leaving
us with Quantum Mechanics on the Higgs branch as the DLCQ (since we
are in 0+1 or 1+1 dimensions, there are large fluctuations in the
ground state and the Higgs branch or Coulomb branch are not really
moduli spaces of the theory. Therefore decoupling is actually not
automatic as we portrayed it. For the different arguments for the two
cases the reader is referred to the literature).

Let us be a little more precise. In the limit
$g_{ym}^2\rightarrow\infty$ three things happen:

\item{1.} The $F$ and $D$ constraints are imposed exactly, giving us 
a non-linear sigma model.

\item{2.} If we are on the Higgs branch, and do not rescale the 
Higgs variables as we take $g_{ym}^2\rightarrow \infty$ then the
following happens. To keep $Y_1^2Q^2+Y_1\mu^2$ fixed we also do not
perform any $g_{ym}$ dependent rescaling on $Y$. The result is that
the kinetic term of the vector multiplet vanishes. These fields become
auxiliary fields and we can integrate them out. This, for example,
gives the 4-Fermi interaction in the non-linear sigma model.  The
value of $Y_1$ in terms of the Higgs branch variables is determined by
the coupling of $Y_1$ in the Lagrangian and is therefore given
(qualitatively)
\eqn\defu{{1\over 2}\{Y_1, QQ^\dagger + {\tilde Q}^*{\tilde
Q}^T\} +[H,[H,Y_1]]=\ fermion\ bi-linear} (The details of the
formula change from dimension to dimension since the number of $Y_1$
variables changes)

\item{3.} Finally, the physical Hilbert space is restricted to $U(N_0)$ 
invariant wave functions.

The main point of this paper is the following. It would seem that we
have lost all information about the Coulomb branch. We will argue that
this is not the case. The scaling that we have performed gave a finite
value to the coordinates $Y_1$ as a function of the Higgs branch
variables. We have used this fact to integrate out the $Y_1$
variables. But we can also look at things in a different way and
regard the $Y_1$'s, which prior to decoupling were coordinates on the
Coulomb branch, as operators on the Higgs branch. Suppose we are given
a time varying quantum state on the Higgs branch $\wvfnc{\Psi(t)}$,
then by computing
\eqn\newvar{Y(t)=<\Psi(t)\vert Y(Q,H,\mu,\theta')\vert \Psi(t)>}
we can transform the problem to that of an excitation moving on the
Coulomb branch. The operation is nothing but a change of basis to a
basis of eigenfunctions of the $Y_1$ operators.

What we will show is that given a Higgs branch sigma model as a DLCQ
of some decoupled high dimension quantum mechanical system, then the
$Y_1$ coordinates describe a DLCQ version of the near horizon of
limit of this system. Hence the Maldacena conjecture for these cases
is essentially a change of basis in Matrix theory.

This will be the main point of the paper, but it is not as
straightforward as we have made it to be. The reason is that the $Y_1$
operators do not commute, and therefore we can not go to a basis of
wave functions which are eigenfunctions of all $Y_1$'s and describe
the system as a particle moving on a new commutative space. Rather the
new space will be non-commutative. We will show, however, that in the
limit in which the near horizon geometry becomes flat, the
non-commutativity disappears. This aspect of the near-horizon limit is
similar to ideas put forward in \ansanj, although the precise relation
remains to be clarified.

The construction discussed here is also similar to that of
\insta. 
There also coordinates analogous to the Coulomb branch reappear and
become the coordinates which complete the 4 coordinates of the
D3-brane to 10 coordinates of the bulk theory. There is however a
difference in the sense that here we are discussing the exact quantum
mechanical description in spacetime whereas the computation there
applies to leading terms only in instanton computations. More closely
related to the construction there might be an IKKT type conjecture
\ikkt. The conjecture might be that the exact finite curvature
$AdS_5\times S^5$ has an IKKT-like description in terms of the large
$k$ limit of $k$ instantons in $SU(N)$, even for finite $N$. Even
though this is conjecture is natural, it is somewhat disturbing. To
see that, let us consider how this conjecture might be proven. The
simplest way to relate the IKKT conjecture to the $D=4, {\cal N}=4$
field theory is if in the large t'hooft coupling the full field
theoretic path integral is approximated by instanton configuration,
and their vicinity. This is not unlikely, as the action of other
configurations is not protected and it can receive strong corrections
and become infinity, leaving us only with approximate instanton
configurations. The problem with this is that for small t'hooft
coupling (i.e., large curvatures) an IKKT conjecture is implausible,
because in that case the path integral clearly has other
contributions.

\subsec{The Matrix description of the (2,0) CFT}

The Matrix model for the 6 dimensional (2,0) Superconformal field
theory was discussed in \twozera,\twozerb. In this section we will
briefly revisit it, in view of the construction from the previous
section. In this case we would like to make some preliminary contact
with the near horizon limit of this theory \juan.

The $N_0$ units of momenta Matrix model for $N_4$ longitudinal 5-branes
coupled to 11D supergravity is discussed in \md\twozera. The
Lagrangian of the Quantum Mechanics is

\eqn\symzd{  \int dt 
 {1\over R}(\partial Y_{-1})^2 + 
\theta(\partial\theta) +RM_p^3\theta [Y_{-1},\theta] + 
RM_p^6[Y_{-1},Y_{-1}]^2}
$$ + {1\over R}(\partial H)^2 + RM_p^6[Y_{-1},H]^2 + 
\theta'(\partial\theta') +M_p^3\theta'[Y_{-1},\theta']  + $$
$$ + {1\over R}(\partial Q)^2 + RM_p^6 Y_{-1}^2Q^2 + RM_p^3\mu(Y_1\mu) +
     RM_p^6 ([H,H]+Q^2)^2 $$
The Lagrangians \symzd\ and \symd\ are related by setting
\eqn\gcoup{g_{ym}^2=R^3M_p^6} and by rescaling of fields. In particular
\eqn\rescy{Y_1=RM_p^3 Y_{-1}.}

$Y_{-1}$ is the distance transverse to the brane (in the canonical
metric, before we took into account the back-reaction of the
brane). In the near horizon limit \juan\ of the (2,0) field theory we
rescale this coordinate such that a dimension 2 combination
$$Y_{nh}=M_p^3 Y_{-1}$$ remains finite. We see that the relation is
\eqn\nhlm{Y_{nh}={1\over R}Y_1,} which implies that a finite $Y_1$ 
coordinate is equivalent to a finite $Y_{nh}$. Therefore, in the SYM
decoupling limit it is precisely the near horizon coordinate which
remains finite and is given as an operator on the Higgs branch ($R$ is
kept fixed throughout).

Another perspective on DLCQ and the AdS/CFT for the (2,0) field theory
is given in \ah.

\subsec{The Matrix description of the ``little string theory''}

The Matrix description of the 6 dimensional ``little string theory''
\dvv\brs\natidec\ was discussed in \twozera\ and \edltls. Let us begin 
with the model before we go the ``little string theory'' decoupling
limit
\natidec. The model is the 1+1 dimensional model discussed above, 
on a cylindrical worldsheet with radius
\eqn\rad{\Sigma_1={1\over R{\tilde M}_s^2}} where ${\tilde M}_s$ is the 
mass scale associated with the ``little string theory'', and the SYM
coupling is
\eqn\ltlgcp{{1\over g_{ym}^2}= {{\tilde M}_s^2\over R^2M_p^6}}
where $M_p$ is the 11 dimensional Planck scale which is taken to
infinity, keeping ${\tilde M_s}$ and $R$ fixed.

As before the dimension 1 $Y_1$ coordinate is fixed in terms of the
Higgs branch variables as we go to the decoupling limit. Again, this
coordinate is related to the dimension -1 coordinate by
\eqn\ltlnhb{Y_{-1}={Y_1\over RM_p^3}.}
 This is exactly the correct scaling which is required to focus on the
near horizon limit of the supergravity dual of the ``little string
theory'' \ltlnh\ (i.e., the near horizon limit of the CHS vacuum
\chs). Again we see that the finite $Y_1$ coordinate is precisely the
quantity which focuses on the near-horizon limit.

We will devote the rest of this paper to the D0-D4 case. It is easy
for formulate, however, the expected results for the D1-D5 as
well. One expects that (in the RR sector) the $Y$'s as functions of
the Higgs branch variables will define operators that will be the
coordinates of the DLCQ of the near horizon limit of the NS 5-brane -
i.e., the linear dilaton background capped by $AdS_7\times S^4$
\ltlnh.  In particular, to generate weakly coupled string perturbation 
theory in the linear dilaton region one needs to go to the long string  
in the sense of 
\lngstrng. The relevant winding number will now be given by the 
winding of the $Y_1$ composite operator.

The reappearance of the Coulomb branch is particularly interesting for
the D1-D5 in view of \natied\ (The system there is not precisely the
one we are discussing here there but is clearly related. There it is
the instanton moduli space on $T^4$ (which does not have a finite
dimensional hyperk\"ahler quotient construction) and here we are
discussing the instanton moduli space on $R^4$). It was shown there
that associated with the singularities in the sigma model target space
there is a new continuum of states in the CFT, suggesting a kind of
tube. We see here that there is a natural way of regenerating this
tube (although the precise relation is not clear).

We have specialized to the cases of the (2,0) field theory and of the
``little string theory''. There are other cases, however, in which
theories are given by the decoupling of Higgs and Coulomb branches, and
we expect that our approach will be just as valid there. Most other
examples, such as 4D ${\cal N}=4$, are given in terms of an impurity
system (for example \orisav). The procedure outlined above should work
just as well, except that we will also need to use DLCQ open string
field theory (this will be the general case for theories on
D-branes). This issue will be discussed in \forthcoming.


\newsec{More on the map}

We have seen that the near horizon Coulomb coordinate can be thought
of as an operator on the Higgs branch sigma model. This operator is
determined by the equations of motion of $Y$. The equation of $Y$ is of
the form
$${\bf L}Y^{[AB]}=\Lambda^{[AB]}$$
where $\Lambda$ is a fermion bi-linear and ${\bf L}$ is
$${\bf L}Y = {1\over 2}\{Y, QQ^\dagger + {\tilde Q}^*{\tilde
Q}^T\} +[H_n,[H_n,Y]].$$ We therefore need to discuss the
invertibility of ${\bf L}$. 

Following \insta\instb, it is straightforward to show that the
operator is non-negative, on the space of Hermitian matrices. This
can be seen by the fact that
\eqn\opnrm{Tr\bigl(\Omega^\dagger {\bf L} \Omega\bigr)=
{\vert \Omega Q\vert}^2+{\vert \Omega {\tilde Q}^*\vert}^2+ {\vert
[H_n,\Omega]\vert}^2.} Using the equation it is also easy to see that
${\bf L}$ is not invertible only at singular points on the ADHM moduli
space, i.e., points where parts of the $U(N_0)$ is restored. The
reason is that if ${\bf L}$ has a zero eigenvalue, then the
corresponding (perhaps a few) eigenvectors $\Omega$ are Hermitian
matrices such that $exp(i\Omega)$ is a subgroup of $U(N_0)$ which
leaves $Q,{\tilde Q}$ and $H_n$ invariant.

Hence at generic points of the moduli space, the map is invertible. It
will be interesting to study further the behavior of the map as one
approaches one of the singular points. For example, some of these
singularities correspond to instantons shrinking to zero size. As we
approach such a point, the inverted operator ${\bf L}^{-1}$ will
diverge which corresponds to at least one eigenvalue in the $Y$
coordinate going to $\infty$. This is to be expected from the UV/IR
relation
\uvir. Since this describes a point like object in the field theory,
it will be associated with an excitation of the bulk at $Y$
approaching $\infty$.

\subsec{The UV/IR relation and the origin of the Higgs branch}

Let us elaborate somewhat the relation between \defu\ and the UV/IR
relation \uvir. Although the following discussion follows from
conformal invariance, and thus not really a test of the map, it does
demonstrate some interesting aspects of the map. A related discussion
appears in \mp.

The $SO(5)$ quantum numbers of a state can only be carried by the
fermion bi-linear terms on the RHS (since bosons on the LHS do not
carry $SO(5)$ quantum numbers). Suppose we fix these quantum numbers,
i.e., we fix the RHS (in the sense of its action on a state). Now
consider making the state smaller and smaller in 4 remaining
coordinates of the brane (transverse to time and to the null circle
directions). This means that the values of the LHS get smaller and
smaller. When we solve for $Y_1$, i.e., the position of the excitation
in $AdS_7$ (with fixed the quantum numbers on the $S^4$), then it
increases. I.e., as the object becomes more and more localized, its
image in $AdS_7$ approaches the boundary $Y\rightarrow\infty$. 

The map \defu\ in effect seems to regulate the singularity at the
origin of the Higgs branch (neglecting the decoupled center of mass
coordinate) since it maps it to the boundary of AdS, which is regular.
Furthermore, from this point of view it is natural to have the 6D
tensor multiplet of the M5-branes at the boundary of $AdS$, since in
the Higgs branch picture one can locate the corresponding state at the
origin of the Higgs branch. `Natural' in this context means that, if
we would blow up the singularity, this state mixes with all the rest
\twozerb.


\newsec{The Higgs branch for a single D0 brane}

The D0-D4 system is easier to analyze and we will focus on it. We have
seen that for all $N_4$ and $N_0$ there are operators on the Higgs
branch which are the coordinates on the near horizon space-time.  In
the remainder of this paper we will focus on the case of $N_0=1$. Even
though the Matrix theory limit requires $N_0\rightarrow\infty$ for
fixed $N_4$, this simple case demonstrates some important aspects of
the construction\foot{It is actually particularly interesting to
better understand the map for $N_0>1$ in the limit that both $N_0$ and
$N_4$ are taken to infinity. This limit might be useful in
understanding the structure of the bound state of (only) the
D0-branes. We are able to map the Higgs branch to dynamics on the
Coulomb branch with arbitrarily low curvature (set by $N_4$). In
particular, there exist states on the Higgs branch which correspond to
the bound states of D0-branes in the near horizon Coulomb branch. This
map, therefore, maps a rather complicated non-abelian problem to a
problem in a sigma model with no non-abelian gauge dynamics. For
example, proving the existence of these states on the Higgs branch is
simpler then proving their existence in flat space Matrix theory
\bndstt. In fact, the counting can be easily done for any $N_4$ and
$N_0$ \nak\ (and adapted for Matrix theory purposes in
\twozerb). As we take $N_0\rightarrow\infty$ the bound state on the
near horizon Coulomb branch expands in the Coulomb branch
coordinates. To approach the flat space limit we also need to take
$N_4\rightarrow \infty$, such that the size of the bound state is always
smaller then the radius of curvature of $AdS$. In this case we can
regard the wave functions on the sigma model as a regulated version of
the bound state in flat space, and we can remove the regulator at
will.}.

\subsec{Quantum mechanics on the ADHM moduli space}

The field
content of the ADHM moduli space for $N_0=1$ and arbitrary $N_4$ is
the following:
\eqn\representations{\matrix{
   &      U(1)_{gauge}& SU(2)_R& SU(2)_L& Spin(5)& U(N_4)\cr
H  &      0           & \bf{2} & \bf{2} & \bf{1}&  \bf{1}\cr
\Theta&   0           & \bf{1} & \bf{2} & \bf{4}&  \bf{1}\cr
w  &     \bf{1}       & \bf{2} & \bf{1} & \bf{1}&  \bf{N_4}\cr
\mu     &\bf{1}       & \bf{1} & \bf{1} & \bf{4}&  \bf{N_4},\cr}}
The $H-\Theta$ multiplet is free, and for the most part we wil ignore
it (an identical sector will appear in the near horizon Coulomb branch
and the identification between them is immediate).  Generally,
$\alpha,\beta..$ will denote $SU(2)_R$, $i,j..$ will denote $U(N_4)$
indices and $A,B,..$ will denote spinor $Spin(5)$ indices ($w$ is what
we called $Q$ before).

The hypermultiplets $w$ satisfy D-term constraints
\eqn\dcnst{ {\bigl(\sigma^a\bigr)}^\alpha_\beta 
w_\alpha^j {\bar w}^\beta_j=0,\ a=1..3 } and $j$ is the $U(N_4)$ index
(whenever possible we will use the notation of \insta). These
constraints imply that $w_\alpha^j {\bar w}^\beta_j$ is proportional
to the identity $2\times 2$ matrix, and it is convenient to define
\eqn\defq{Q^2={1\over2}Tr(w_\alpha^j {\bar w}^\beta_j)}
The fermions are also restricted by the relation
\eqn\frmcnst{ {\bar \mu}^A_j w_\alpha^j + 
J^{AB}\epsilon_{\alpha\beta} \mu_B^j {\bar w}^\beta_j=0.}  This
equation comes from integrating out the superpartners of the Coulomb
branch coordinates ($J^{AB}$ denotes the antisymmetric form of
$USp(2)$). In addition we need to mod out by $U(1)_{gauge}$.

Although we will discuss in a moment the solution to these equations
when we fix the $U(N_4)$ symmetry, let us first proceed more
generally. As in \insta, it is convenient to classify the solutions to
the linear constraint \frmcnst\ in the following way. One class of
solutions, which we will denote by $\nu^B_k,\ k=1..N_4-2$,
satisfies that $\mu$ and ${\bar w}$ are orthogonal (in flavor indices,
for every $A$ and $\alpha$). The second class of solutions is given by
\eqn\etasol{\matrix{
\mu^j_B       ={1\over Q}w^j_\alpha \eta^\alpha_B\cr
{\bar \mu}^B_j={1\over Q}{\bar\eta}_\alpha^B {\bar w}^\alpha_j}}
where the $\eta$ satisfy the pseudo-reality condition 
\eqn\psdrlty{
{\bar\eta}^A_\alpha +J^{AB}\epsilon_{\alpha\beta}\eta^\beta_B=0.}
Regarding this Higgs branch as the moduli space of instantons these
are nothing but the superconformal zero modes.

\subsec{The reduced model}

We are interested in mapping states in this quantum mechanics to
states on the near horizon Coulomb branch. There is, however, an
obstacle which is that many states on the Higgs branch carry $U(N_4)$
quantum numbers, which clearly does not exist on the Coulomb branch
side. We will therefore restrict ourselves to states that are
invariant under $U(N_4)$. This is compatible with \defu, which is also
$U(N_4)$ invariant\foot{ The interpretation of states that are
charged under this group is not clear.}.
 
To discuss states that are invariant under $U(N_4)$ (and under
$U(1)_{gauge}$), it is convenient to define a ``reduced'' quantum
mechanics, which is the quantum mechanics on a Hilbert space of flavor
invariant states (as well as gauge invariant, of course). Actually we
will fix the gauge only partially and work with a larger Hilbert
space, to which we will also refer at times as the ``reduced'' Hilbert
space. The procedure that we will adopt is to gauge fix the bosonic
part of the hypermultiplets. A convenient choice of gauge will be
\eqn\qgg{w_\alpha^j=\biggl(\matrix{ Q & 0 & 0 & .. & 0\cr
				    0 & Q & 0 & .. & 0\cr}\biggr).} It
is clear that gauge and flavor invariant functions can be thought of
as functions of a single variable $Q$, but with as many fermions as we
had before, i.e., this space has a single bosonic coordinate, 8
fermionic coordinates $\eta$ and $4(N_4-2)$ complex pairs of fermions
$\nu^A_k,{\bar\nu}^k_A$. We have clearly not fixed the gauge
completely and, for example, still have $U(N_4-2)$ flavor symmetry
acting on the fermions which we will now eliminate

The commutation relation on the remaining fermions can still be taken
to be the canonical ones:
\eqn\comrel{
\{\eta^\alpha_A,\eta^\beta_B\}=\epsilon^{\alpha\beta}J_{AB},\ \
\{\nu^A_k,{\bar\nu}^{k'}_B   \}=\delta^k_{k'}\delta^A_B }

\medskip

\noindent{\it The reduced Hilbert space}

\medskip

We start with a model that has $SU(N_4)_{global}\times U(1)_{gauge}$
symmetry. The $U(1)_{gauge}$ is the diagonal of the $U(N_4)$ that acts
on the hypermultiplets. After we go to the special gauge above the
remaining symmetry is $SU(N_4-2)_{flavor}\times U(1)_{gauge}$. We are
interested in states which are invariant under all these
symmetries. Clearly $Q$ and $\eta$ are invariant under these
symmetries, hence all the restrictions will be in the $\nu-{\bar\nu}$
Fock space. There are two states in this Fock space,
$\wvfnc{+}$ and $\wvfnc{-}$, which satisfy
\eqn\fcstt{\nu^A_k     \wvfnc{-}=0,\ \ \ 
          {\bar\nu}^k_A\wvfnc{+}=0.} These states are exchanged under
$\nu\leftrightarrow\bar\nu$, hence their natural $U(1)$ charge
assignment will be opposite. If we set the $U(1)$ gauge charge of
$\nu$ to 1, then $\wvfnc{+}$ has charge $-2(N_4-2)$. The requirement
of gauge invariance then tells us that we are restricted to states of
the form $\nu^{2(N_4-2)}\wvfnc{+}$. Furthermore, since all the indices
on the $\nu$ operators are that of a fundamental index, we can only
contract them by a baryonic vertex. The Hilbert space of gauge
invariant and flavor invariant functions is therefore of the form
\eqn\hlbrta{
f(Q,\eta)\eps^{l_1..l_{N_4-2}}\eps^{t_1..t_{N_4-2}}
\nu^{A_1}_{l_1}..\nu^{A_{N_4-2}}_{l_{N_4-2}}
\nu^{B_1}_{t_1}..\nu^{B_{N_4-2}}_{t_{N_4-2}}
\wvfnc{+}}
for brevity we will denote $N_4-2$ by $N$.

\medskip

\noindent{\it Aspects of the reduced Supercharges and Hamiltonian} 

\medskip

We have seen that the structure of the Hilbert space is simple
enough. One would like to know whether ``reducing'' in this way has
not made the dynamics, as encoded by the Hamiltonian or supercharges,
too complicated. Fortunately, The price that we pay for reducing is
minimal. In the full ADHM quantum mechanics flavor invariance is the
equation
\eqn\grpact{(T^a_{bos}+T^a_{ferm})\Psi(w,{\bar w},\mu,{\bar\mu})=0,} 
where $a$ is an $u(N)$ index, $T^a_{bos}$ is the action on the bosons
and $T^a_{fer}$ is the action on the fermions. \grpact\ equates
derivatives with respect to $w,{\bar w}$ (along orbits of the flavor
symmetry) with fermionic bi-linear operators acting on the states. We
can now replace the bosonic derivatives in these directions with
fermion bi-linear operators. The only bosonic derivative that will be
left is in the $Q$ direction. We have, however, generated new fermion
tri-linear terms, but since the original supercharges also contained
tri-linear fermion terms, this does not complicate the system (A toy
example of this reduction procedure is discussed in appendix 1).

Given the supercharges of the initial ADHM quantum mechanics we can
determine the supercharges of the reduced system. Alternatively since
we know that the supercharges have at most tri-linear fermion terms we
can compute them explicitly by requiring closure of the supersymmetry
algebra. The supercharges are computed in this way in appendix 2, and
the result is
\eqn\sprcrg{Q^\alpha_A=\eta^\alpha_A
\biggl({\partial\over\partial Q}+{a\over Q}\biggr)+
{2\over Q}\et{\al}{B}L^{BC}J_{CA}+ {1\over
Q}M^{\al\be}\et{\ga}{A}\epsilon_{\be\ga}.}  where $L$ is the $SO(5)$
generator in the $\nu-{\bar\nu}$ sector, $M$ is the $SU(2)_R$
generator on $\eta$ (for precise conventions see appendix 2), and $a$
is a specific constant which we have not determined since will we not
require it.


\newsec{DLCQ of the Poincare patch of $AdS_7\times S^4$}

We would like to perform a DLCQ quantization of M-theory on
$AdS_7\times S^4$.  The correct and full DLCQ of this background is of
course the quantum mechanics on the ADHM moduli space. We would like,
however, to start with this spacetime and try and write an approximate
DLCQ description of that.  This step, however, is somewhat problematic. 
We can not try and
write a DLCQ for the entire full cover, because this space does not
have any null isometries.  However, the Poincare patch has a null
isometry.  If we use the coordinates $U,X^i,\ i=0..5$ then we can mod
out by the symmetry $x^+\rightarrow x^++R$. The problem now is that,
since there is a fixed point, the quotient is singular at $U=0$.  Away
from the singularity we can write a DLCQ, which will be a finite
dimensional quantum mechanics. As we approach the fixed point of the
null translation, the quantum mechanics will become singular. This may
invalidate the whole approach but we will argue that this is not the
case.  Having a singularity in the Hamiltonian of the DLCQ is not a
source for concern as long as one asks the right questions.  There is,
for example, no point in asking what is the ground state of the
quantum mechanics, but if the Hamiltonian is a good differential
operator on functions supported away from the singularity,
then it is a sensible question to ask what is the dynamics of a wave
packet there.

We will restrict our attention to the DLCQ for $N_0=1$, i.e., for a
single unit of momentum along the null circle. In this case the model
is a quantum mechanical sigma model, which describes the motion of the
gravity multiplet on this background. Taking the metric of
$AdS_7\times S^4$ to be\foot{We use the coordinate $r$ which is
proportional to the distance from the brane in the uncorrected metric}
$$rdx^2+{\bigl( {dr\over r}\bigr)}^2 + d\theta^2$$ the
equation of motion of a scalar particle of mass $m$ is
$$\biggl( \partial_x^2 + 
{1\over r}\partial_r r^4 \partial_r 
+r\partial_\theta^2 + rm^2 \biggr)\Psi=0.$$
Going to DLCQ
$$H\sim{1\over P_+} \biggl(
\partial_{x_\perp}^2 + 
{1\over r}\partial_r r^4 \partial_r 
+r\partial_\theta^2 + rm^2 \biggr)$$
which is a the Hamiltonian on a sigma model with metric
$$dx^2+{1\over r^3}(dr^2 + r^2 d\theta^2)$$ up to a shift in the mass
and an $r$-dependent rescaling on $\Psi$.  Not surprisingly, this is
the same as the metric that a D0 sees near a D4 brane, when we rescale
the coordinates as we go to the M5-near horizon limit.

The full Lagrangian of a D0 away from a D4 brane was determined, to
lowest order in derivative expansion, in \raem. In addition to a
decoupled $R^4$ (associated with the hypermultiplet $H$ which is
decoupled for $N_0=1$) there are 5 coordinates $U^i$ ($U^2=\Sigma
{U^i}^2$) which\foot{Generally the letter $U$ will denote commuting
coordinates on the Coulomb branch. $Y^i$ denotes the non-commuting
ones.} parameterize 5 coordinates transverse to the brane, i.e, one
radial coordinate and the four of $S^4$. The most general form of an
$SO(5)$ invariant Lagrangian is given in equation (3.10) in
\raem, and it is:
\eqn\adsdlcq{
-f(U)\bigl( {\dot U}^j{\dot U}^j + i(
{\bar\rho}{\dot\rho}+\rho{\dot{\bar\rho}} ) \bigr) +{\dot
U}^if_{,j}(\rho\gamma^{ij}{\bar\rho})+ {1\over2}
\bigl(f_{ij}-f^{-1}f_{,i}f_{,j}\bigr) (\rho\gamma^i{\bar\rho}
\rho\gamma^i{\bar\rho} +\rho\gamma^i\rho {\bar\rho}\gamma^i{\bar\rho})
} where $\rho$ (which is denoted in \raem\ by $\eta$) is a ${\bf 4}$
of $USp(2)$ and $\rho$ and ${\bar\rho}$ together form a doublet of
$SU(2)_R$.  The most general $f(U)$ allowed by the (super)symmetries
is $f=c_0+{c_1\over U^3}$.  This is the metric that a D0 brane sees
near a D4-brane (for proper $c_{0,1}$). When we take $M_p\rightarrow
\infty$, and properly normalize the $U$ coordinates as to go to the 
near horizon limit, we obtain that the function $f$ is given by
\eqn\newf{f(U)={N_4\over RM_p^3 U^3}.}

In section 6.5 we will partially match terms in the supercharges for
this system with terms in the supercharges of the ADHM sigma
model. The term which is most interesting to match is the one that
contains derivatives with respect to $U^i$ because transforming this
term into a Higgs branch expression will rely most heavily on the maps
from the Higgs to the Coulomb branch. The $\partial_U$ term appears in
the Coulomb branch supercharges as
\eqn\hgscrg{Q^\alpha_A=U^{3\over2}\rho^\alpha_B J^{BC}\gamma^i_{AC}
{\partial\over\partial U^i}+\ 3\ \rho\ terms}

One more point is in order. We have restricted ourselves to
$N_0=1$. One can ask whether there is a generalization of this formula
to larger values of $N_0$.  There are several ways to try and obtain
approximate answers but, since there is no established supersymmetric
non-renormalizaton theorem for these cases, we will not explore this
generalization here. One thing is clear, however: when the quanta that 
carry DLCQ momenta are close to each other (compared to the scale set by
the curvature of space-time) we should approximately obtain the ${\cal
N}=16$ BFSS Lagrangian as a limit of the ADHM sigma model.


\newsec{Higgs Coulomb equivalence}

\subsec{The coordinate algebra}

The relation \defu\ defines a set of operators $Y^{[AB]}$ which are
coordinates on the near horizon Coulomb branch as operators on the
Higgs branch. This relation is valid for all $N_0$ and $N_4$. For the
case $N_0=1$ this relation simplifies significantly and becomes
\eqn\nzrrl{Y^{[AB]}=
{{\nu^A_k{\bar\nu}^k_FJ^{BF}-\nu^B_k{\bar\nu}^k_CJ^{AC}
+{1\over2}J^{AB}\nu^F_k{\bar\nu}^k_F}\over Q^2}} where the sum over
$k$ is $k=1..N$ ($N=N_4-2$).  Using this formula we would like to
study in greater detail how the correspondence works.
 
The equation \nzrrl\ defines the operators $Y$ in terms of $Q^2$ and
in terms of the fundamental fermions. In the context of $AdS_7\times
S^4$ we should think of the 5 $Y$ coordinates as $R^+\times S^4$,
where the $R^+$, which we will denote as $\bar U$, is associated with
the additional coordinate of $AdS_7$. We will see below that
\eqn\urela{{\bar U}\sim {\sqrt{2} (N)\over Q^2},} and the $S^4$
manifold (which becomes fuzzy) is related to the fermion bi-linear. We
therefore begin by focusing on the fermion bi-linear terms.

Before doing so, it is useful to list the main objects that we will be
dealing with, which is the following ``algebra of coordinates'', and
some of the relations between them:

\medskip

\noindent 1. The flavor invariant coordinates:
\eqn\eqbi{B^{[AB]}=
\nu^A_k \bnu{}_F^k J^{BF}- 
\nu^B_k \bnu{}_F^k J^{AF}+
{1\over 2}J^{AB}\nu^F_k\bnu{}^k_F }

\noindent 2. The flavor charged coordinates:
\eqn\eqbik{B^{[AB]}_{l_1l_2} = 
\nu^A_{l_1}\nu^B_{l_2}-
\nu^B_{l_1}\nu^A_{l_2}+
{1\over2}J^{[AB]}J_{CD}\nu^C_{l_1}\nu^D_{l_2} }

\noindent 3. The null coordinate
\eqn\eqbz{B^0_{l_1l_2}=J_{AB}\nu^A_{l_1}\nu^B_{l_2}}

$J$ is the antisymmetric forms of $USp(2)$ and we will at times denote
$B^{AB}_{l_1l_2}$ and $B^0_{l_1l_2}$ by $B^{AB}_{..}$ and $B^0_{..}$
respectively.

The operators $B^{AB}$ and $B^{AB}_{..}$ are a ${\bf 5}$ of $USp(2)$,
and $B^0_{..}$ is a singlet. The $B^{AB}$ operators satisfy the
following hermiticity relation
\eqn\hrmrel{ { {\bigl(B^{AB}\bigr)}^\dagger}_{AB}=
{B^\dagger}_{BA}=
J_{AA'}B^{A'B'}J_{B'B}} and are 
therefore related, as expected, to 5 real coordinates.
The relation is
\eqn\fvrl{B^{13}=B^1+iB^2=-{B^{42}}^\dagger,\ 
          B^{14}=B^3+iB^4= {B^{32}}^\dagger,\ 
          B^{12}=B^5=-B^{34}.}
\eqn\jja{ J_{AC}J_{BD}B^{AB}B^{CD}=4\Sigma_{i=1}^5 {B^i}^2}

Some of the relations between these operators, which will be useful
below, are:

\eqn\bicmm{ [B^{[AB]}, B^{[CD]}]=
J^{AC}L^{BD}-J^{AD}L^{BC}-J^{BC}L^{AD}+J^{BD}L^{AC}.}

\eqn\algb{ [B^{[AB]},B^0_{l_1l_2}]=2B^{[AB]}_{l_1l_2} }

\eqn\algc{ [B^{[AB]},B^{[CD]}_{l_1l_2}]=
\biggl(J^{AC}J^{BD}-J^{AD}J^{BC}-{1\over2}J^{AB}J^{CD}\biggr)
B^0_{l_1l_2} }

\eqn\algd{ J_{AC}J_{BD}B^{[AB]}_{l_1k_1}B^{[CD]}_{l_2k_2} = }
$$\bigl( {1\over2}B^0_{l_1k_1}B^0_{l_2k_2} - (l_1\leftrightarrow l_2)
- (k_1\leftrightarrow k_2) + (k_1,l_1\leftrightarrow k_2,l_2) \bigr) +
... $$ where $...$ are terms which are symmetric under $l_1\rightarrow
l_2$ or $k_1\rightarrow k_2$ and, as will be clear below, will not
play a role in our analysis. These operators and the relations between
them are the structures which organize the correspondence for the
$N_0=1$ case.

Note equation \bicmm\ which encodes the non-commutativity of the
coordinates. The non-commutativity is closely related to the
truncation of the spectrum of KK states on $S^4$, and since in the
ADHM quantum mechanics the spectrum of chiral operators truncates
correctly for every $N_4$ and $N_0$ we expect that this
non-commutativity will persist (in some form) even in this limit. This
implies that spacetime becomes ``non-commutative'' even for closed
strings in the presence of a closed string field strength. The length
scale associated with this non-commutativity may be much smaller than
the string scale, but since D-branes can probe sub-stringy structures,
they can still resolve it, which is another interpretation of our
results for the D0-D4 system.

\subsec{Mapping of states}

\noindent{\it The Fermi surface}

\medskip

The purpose of this section is to show how the operators of the
coordinate algebra organize the states of the reduced system. We will
show that the 5 $B^{AB}$ operators organize these states in the same
way that the  5 commutative coordinates of $R^5$
organize the spherical harmonics on the unit sphere.

The class of relevant states is given in \hlbrta. Each baryonic 
contraction $\eps^{....}\nu^{A_1}_{.}..\nu^{A_N}_{.}$ is the
N-th tensor symmetric representation of the ${\bf  4}$ of $USp(2)$.
Therefore all the states are in the product of two such 
representations. In particular there is an $USp(2)$  singlet state
given by
\eqn\sttb{\wvfnc{\phi }= 
\epsilon^{l_1..l_{N}}\epsilon^{t_1..t_{N}}
B^0_{l_1t_1}..B^0_{l_{N}t_{N}}\wvfnc{+} } This state corresponds to
the lowest spherical harmonic, i.e., the constant function on the
sphere.

We will refer to this state as the ``Fermi surface''. The reason
is that all higher spherical harmonics will be given as 
excitations of this state, in a similar way to exciting a 
fermion from below a Fermi surface to a level above it.

\medskip

\noindent {\it Higher spherical harmonics}

\medskip

The rest of the spherical harmonics on $S^4$ match with states in the
Fock space by the following correspondence. Suppose we are given a
j-th symmetric traceless tensor of $SO(5)$ which we will write in
$USp(2)$ conventions as $V_{[A_1B_1]..[A_jB_j]}$, then the map \nzrrl\
implies
\eqn\spmtch{ V_{[A_1B_1]..[A_jB_j]}U^{[A_1B_1]}...U^{[A_jB_j]} 
\leftrightarrow
V_{[A_1B_1]..[A_jB_j]} B^{[A_1B_1]}...B^{[A_jB_j]}
\wvfnc{\Phi}=}
$$=2^j{N!\over (N-j)!} V_{[A_1B_1]..[A_jB_j]}
\eps^{...}\eps^{...}B^{A_1B_1}_{..}... B^{A_jB_j}_{..}
B^0_{..}...B^0_{..}\vert +> .$$

We have restricted the $V$'s to live in a single irreducible
representation. If we allow general $V$ then the relation is still
correct as long as we do not use on the LHS the relation $U^2=Const$,
but leave it as an operator $U^2$. The reason is that $B^2$ evaluated
on the different wave function on the RHS is not a constant (although
for large N the corrections are roughly suppressed by powers of
$j\over N$ - this will be discussed further in section 6.4). If we
would allow a general $V$ above and use $U^2=Const$ then each
representation would be over-defined (by all $V$'s with a larger number
of vector indices), and the inequality of $B^2$ on the different
representations would make the definitions incompatible.

\medskip

\noindent {\it Counting of states}

\medskip
 
We would like to show that all the states \hlbrta\ are in 1-1
correspondence with spherical harmonics, and that the spectrum
truncates at the correct place. The last statement is easy to verify
and should comes as no surprise since the ADHM quantum mechanics gives
the correct spectrum of chiral operators \twozerb.  The state with the
largest number of flavor charged coordinates has $N$ such coordinates
$$\eps^{....}\eps^{....}
B^{A_1B_1}_{..}.. B^{A_{N}B_{N}}_{..}
{\vert +>}$$ so there are overall $N_4-1$ spherical
harmonics functions in the spectrum. This exactly matches the expected
truncation of states on the $S^4$ from the $(2,0)$ CFT point of view
(a truncation similar to the truncation to $TrX^2,.., Tr X^{N_3}$ for
$AdS_5\times S^5$).

Supergravity is, in a sense, given by the quantum mechanics 
of the system \adsdlcq. Indeed there we do not see any truncation 
of the spectrum. However, the full non-perturbative formulation,
which is the ADHM QM, includes the correct cut-off.

Next we would like to show that all the states in \hlbrta\ are
generated by acting with the coordinate algebra on the Fermi
surface. We regard the states in \hlbrta\ as living in the product of
two $N$-th symmetric tensor products of ${\bf 4}$'s of $USp(2)$. To
examine what representation appears in this product, it's enough to
choose a fixed vector of our choice in one of the representations. We
will chose this state to be such that all the $USp(2)$ indices are
``1''. The relevant class of states is therefore
$$\eps^{l_1..l_N}\eps^{t_1..t_N}
\nu^1_{l_1}..\nu^1_{l_N}
\nu^2_{t_1}..\nu^2_{t_{n_2}}
\nu^3_{t_{n_2+1}}..\nu^3_{t_{n_2+n_3}}
\nu^4_{t_{n_2+n_3+1}}..\nu^4_{t_N}\vert +>$$
It is clear, however, that this state equals
$$\eps^{l_1..l_N}\eps^{t_1..t_N}
B^{12}_{l_1t_1}..B^{12}_{l_{n_2}t_{n_2}}
B^{13}_{l_{n_2+1}t_{n_2+1}}..B^{13}_{l_{n_2n_3}t_{n_2+n_3}}
B^{14}_{l_{n_2+n_3+1}t_{n_2+n_3+1}}..B^{14}_{l_Nt_N}
\vert +>$$ Hence all the
states in \hlbrta\ are generated by the coordinate algebra.

\medskip

Now that we have identified states which correspond to the spherical
harmonics and operators which can play the role of coordinates, we can
examine the correspondence more carefully. We would now like to
address the questions

\item{1.} Is the map \spmtch\ unitary ?
\item{2.} What is the size of the sphere ?
\item{3.} To what extent are the $B$'s coordinates on a sphere ?

The answer that we will obtain is that for low lying Kaluza-Klein
spherical harmonics, the map is unitary, the states correspond to
spherical harmonics on a sphere of radius $\sqrt{2}N$ and the $B$'s
act as coordinates. Fortunately, low-lying Kaluza-Klein states for our
purposes certainly include all states with wavelength larger the
Planck scale.

\subsec{The dilute gas approximation}

This section will touch upon some aspects of the coordinate algebra in
the ``dilute gas approximation''. This approximation is the leading
term for $j<<N$ where $j$ is the level of the spherical harmonics (we
will show that the leading correction is actually $j/ N^\alpha$ where
$\alpha\sim O(1),\ \alpha<1$ will be determined in the next
subsection).

The ``dilute gas approximation'' is the following. For every $j$ the
state is given by
$$\eps^{l_1..l_N}\eps^{t_1..t_N}
B^{A_1B_1}_{l_1t_1}...B^{A_jB_j}_{l_jt_j}
B^0_{l_{j+1}t_{j+1}}..B^0_{l_Nt_N}\vert +>$$ In the limit
$j<<N$ most of the $B$'s are $B^0$. In this approximation whenever a
computation receives a contribution both from $B^{AB}_{lt}$ and from
the $B^0_{lt}$ we will neglect the former, since the contribution from
such term will be proportional to $j$ vs. contributions from the 2nd
term which will be proportional to $N-j$.

Within this approximation we lose the non-commutativity in the
system, and hence the following things happen:

\item{1.}
the map
$$V_{i_1..i_j}U^{i_1}..U^{i_j}\leftrightarrow
  V_{i_1..i_j}B^{i_1}..B^{i_j}\wvfnc{\Phi}$$
$$\leftrightarrow {(2N)}^jV_{i_1..i_j}
\eps^{l_1..l_N}\eps^{t_1..t_N}
B^{i_1}_{l_1t_1}...B^{i_j}_{l_jt_j}
B^0_{l_{j+1}t_{j+1}}..B^0_{l_Nt_N}\vert +>$$ becomes exact
even for $V$'s which are not irreducible.

\item{2.} The operators $B^{AB}$ are now commuting. The action of 
$B^{AB}$ on the states is now
$$B^{AB}\wvfnc{(A_1B_1)..(A_jB_j)}\sim
2N\wvfnc{(AB)(A_1B_1)(A_2B_2)..(A_jB_j)}$$
and correspondingly
$$[B^{AB},B^{CD}]\wvfnc{(A_1B_1)..(A_jB_j)}=0$$

\item{3.} The states move on a sphere of radius $\sqrt{2}N$.
$$J_{AA'}J_{BB'}B^{AB}B^{A'B'}\wvfnc{(A_1B_1)..(A_jB_j)}\sim$$
$$\sim N^2J_{AA'}J_{BB'}\wvfnc{(A_1B_1)(A_2B_2)..(A_jB_j)(AB)(A'B')}=$$
$$=8N^2\wvfnc{(A_1B_1)(A_2B_2)..(A_jB_j)},$$
which implies a fixed radius of the sphere
\eqn\rad{\Sigma_i {B^i}^2=2N^2}
(this is related to but not precisely the radius of the sphere as seen
in supergravity. To compare with that one needs to compute the
Hamiltonian or the supercharges. Although we will briefly discuss the
supercharges below, we will not touch upon this point).

\subsec{An exact computation}

To examine the departures from the dilute gas approximation we would
like to perform an exact computation of things such as the norms,
$<B^2>$ etc. It will be sufficient to do so in a single state in each
irreducible representation. The states that are the most convenient to
use are the states
$$\vert n>={B^{13}}^n\wvfnc{\phi}, \ \ \ <n\vert = <\phi\vert
{B^{24}}^n.$$ As mentioned above, in a 5-vector notation
$B^{13}=B^1+iB^2={B^{24}}^*$. Therefore ${B^{13}}^n$ will generate a
state already in the symmetric, traceless rep. of $SO(5)$.

We would first like to evaluate corrections to the norms of the
states, vs. the norms of a state moving on a commuting $S^4$.  On the
classical geometry side (on the unit sphere) (we will use the
coordinates $U$ here to denote ordinary commuting coordinates on the
unit sphere)
$${\| {U^{13}}^n \|}^2\propto{\Gamma(n+1)\over \Gamma(n+{5\over2})}$$
or in a form which will be more useful
$$ {  {\| {U^{13}}^n \|}^2\over {\| {U^{13}}^{n-1} \|}^2 } = 
{n\over n+{3\over2}}$$
where ${\| {U^{13}}^n \|}^2 = \int d\Omega {\vert U^{13} \vert}^{2n}$, 
whereas on the quantum non-commutative geometry side
$$<n\vert n>= <n-1\vert B^{24} \vert n>=<n-1\vert n(N-n+1) \vert n-1>+$$
$$2(N-n){2^n N!\over (N-n)!}<n-1\vert \eps^{l_1..l_N}\eps^{t_1..t_N}
B^{13}_{l_1t_1}..B^{13}_{l_nt_n}B^{24}_{l_{n+1}t_{n+1}}
B^0_{l_{n+2}t_{n+2}}..B^0_{l_Nt_N}\vert +>$$ The overlap
in the last line is not zero and is partially determined by quantum
numbers
$$<n-1\vert \eps^{l_1..l_N}\eps^{t_1..t_N}
B^{13}_{l_1t_1}..B^{13}_{l_nt_n}B^{24}_{l_{n+1}t_{n+1}}
B^0_{l_{n+2}t_{n+2}}..B^0_{l_Nt_N}\vert +>=$$
$${n\over n+{3\over2}}<n-1\vert\ {1\over4}J_{CC'}J_{DD'}
\eps^{l_1..l_N}\eps^{t_1..t_N}
B^{13}_{l_1t_1}..B^{CD}_{l_nt_n}B^{C'D'}_{l_{n+1}t_{n+1}}
B^0_{l_{n+2}t_{n+2}}..B^0_{l_Nt_N}\vert +>=$$
$$={n\over n+{3\over2}}\cdot{1\over2}\cdot
\eps^{l_1..l_N}\eps^{t_1..t_N}
B^{13}_{l_1t_1}..B^{12}_{l_{n-1}t_{n-1}}
B^0_{l_nt_n}..B^0_{l_Nt_N}\vert +>$$

Hence the recursion relation is
$$<n\vert n>=2N^2{n\over n+{3\over 2}}\biggl(1+O\bigl({n^2\over
N}\bigr)\biggr)<n-1\vert n-1>.$$ As before the leading size of the
sphere is $\Sigma_i {B^i}^2=2N^2$ and the correction to the classical
recursion relation is suppressed by a factor of $n^2/N$. As we compute
the norm using the recursion relation we accumulate errors yielding an
error in the norm of $\vert n>$ of the order of
\eqn\deviat{n^3\over N.}  
We have seen this number before. It is the momentum expansion of
M-theory. The size of the $S^4$ is $R\sim l_p N^{1/3}$ using this
relation the expression \deviat\ becomes $$\biggl({P\over
M_p}\biggr)^3.$$ This is to be expected since for modes for which low
energy supergravity is legitimate we should see only small deviations
from the behavior of wave functions on a fixed sphere.

Next we would like to evaluate the expectation value $B^2$ as seen by
the $n$-th spherical harmonics
$${1\over4}J_{AA'}J_{BB'}B^{AB}B^{A'B'}\vert n>.$$
We will write it as the sum of two terms:
$$B^{12}B^{12}+{1\over2}\bigl(B^{14}N^{32}+B^{32}B^{14}\bigr)\vert
n>=$$
$$=4(N-n)(N-n-1){N!2^n\over (N-n)!}\eps^{....}\eps^{....}{B^{13}_{..}}^n
\bigl( {B^{12}_{..}}^2+B^{14}_{..}B^{32}_{..}\bigr){B^0_{..}}^{N-n-2}
\vert +>$$
$$+3(N-n)\vert n>$$
$${1\over2}\bigl(B^{13}B^{24}+B^{24}B^{13}\bigr)\vert n>=$$
$$=4(N-n)(N-n-1)\eps^{....}\eps^{....}{B_{..}^{13}}^{n+1}B^{24}_{..}
{B^0_{..}}^{N-n-2}\vert + >+$$
$$\bigl( (N-n)+(n+1)(N-n)+n(N-n+1) \bigr) \vert n>$$
The radius of the sphere is therefore given by
$$2(N-n)(N-n-1)+(N-n)+n(N-n+1)+(n+1)(N-n)=2N^2(1+O({n\over N})).$$

\subsec{The definition of $U$}

We have seen that on the relevant states $$\Sigma_i {B^i}^2\sim
2N^2,$$ and what are the corrections to this equation. The full
definition of the 5 coordinates $$Y^{AB}\sim {B^{AB}\over Q^2}$$
implies
\eqn\ytwo{\Sigma_i {Y^i}^2\sim {2N^2\over Q^4}.}

We could leave the matter at that, i.e., have 5 non-commuting
variables, but since spacetime factorizes into a sphere and the
radial coordinate which become part of the $AdS_7$ we will use the
$B$ operators as constrained coordinates on $S^4$ and define an
additional commuting coordinate ${\bar U}$ by 
\eqn\defuagn{{\bar U}={\sqrt{2} N\over Q^2}.} The advantage of this 
definitions is that all the non-commutativity is associated with parts
of space which carry the flux, i.e., the $S^4$, whereas the $AdS_7$ is
a regular commutative sigma model. This enforces the claim that in
some cases there is a fuzziness in spacetime associated with field
non-zero field strengths.

\subsec{Mapping of Supercharges}

We have seen that there is a natural identification of states on the
Higgs branch as states on the near-horizon Coulomb branch. We would
now like to begin addressing the issue of dynamics, i.e., can we see
that the dynamics on the Higgs branch is mapped to the expected
dynamics on the Coulomb branch. To do so we would like to show how the
supercharges of the two systems are related.  In this section we will
partially show how the supercharges on the Coulomb branch are mapped
onto the supercharges of the reduced model, after we rewrite the
Coulomb branch coordinates in term of Higgs branch variables. In the
process we will also need to discuss the transformation laws of the
remaining fermions from the near-horizon Coulomb branch (where we
denoted them by $\rho$) to the Higgs branch (where we denoted them by
$\eta$).

The supersymmetry generators along the Coulomb \hgscrg\ branch contain
the term
\eqn\clta{U^{3\over2}\rho^\alpha_B\gamma^i_{CA}J^{BC}
{\partial\over\partial U^i}} (we will neglect $R,\ M_p$ and $N_4$
dependence). We will show that under the map \nzrrl\ this term becomes
some of the terms in the supercharges of the Higgs branch. 

Replacing the commuting coordinate $U$ by the non-commutative
definition \nzrrl\ is ambiguous because of ordering issues. The
difference should be terms suppressed by powers of $1/N$ and therefore
beyond the scope of the supercharges \hgscrg\ anyhow (which are just
the leading (in derivatives) terms in the Lagrangian). More
practically, we will manipulate the $U$ variables as commuting
variables in the Coulomb branch supercharges for as long as possible.

Inserting
$${\partial\over\partial U^j}={1\over U^2} 
\biggl( U^i{L'}^{ij}+
U^j\bigl(U^i{\partial\over\partial U^i}\bigr)\biggr)$$ where $L'$ are
$SO(5)$ generators $U^i\partial_{U^j}-U^j\partial_{U^i}$. Defining
new fermion operators
\eqn\defl{\eta^\alpha_A=
\rho^\alpha_B\gamma^i_{CA}J^{BC}{U^i\over U},\ \ \ 
\rho^\alpha_B= {1\over2} \eta^\alpha_B J^{BC}\gamma^i_{CA}{U^i\over U} }
where we have used the Fierz relation:
$$\gamma^i{\gamma^j}^*+\gamma^j{\gamma^i}^*=-2\delta^{ij},\ \ 
{\gamma_j}^*=-J\gamma^j J,$$
$\eta$ will be the fermion fields on the Higgs branch.

Using this definition of $\rho$ as a function of $\eta$ we obtain that
this part of the Hamiltonian is
$$\eta^\alpha_A U^{1\over 2}(U^i\partial_{U^i}) + {1\over 2}
U^{-{3\over2}}\eta^\alpha_{B'}J^{B'D}\gamma^j_{DB}J^{BC}\gamma^l_{CA}
U^lU^i{L'}^{ij}.$$

We can now transcribe this expression to the Higgs branch variables:

\noindent{1.} Using the relation $U\sim 1/Q^2$, the first term becomes 
$$\eta^\alpha_A \partial_Q$$ which is one of the terms on the Higgs 
branch.

\noindent{2.} The 2nd expression can be simplified further. The 
expression $\gamma^j_{DB}J^{BC}\gamma^l_{CA}$ can give us two kinds of
terms with different quantum numbers. One which is proportional to
$\delta^{ij}J_{DA}$ which is zero because
$U^lU^i{L'}^{ij}\delta^{jl}=0$ and another which is proportional to
$\gamma^{ij}_{DA}$. The latter is multiplied by
$$U^lU^i{L'}^{ij}-U^jU^i{L'}^{il}$$ where we have anti-symmetrized on
$l\leftrightarrow j$. This, however, equals $U^2 {L'}^{jl}$ on all
spherical harmonics.  The key to the transcription now is to identify
the $SO(5)$ generators $L'$ on the coordinates of $S^4$ as the $SO(5)$
generators $L$ on the $\nu$-fermion (which is the full $SO(5)$
generator on the reduced system). Furthermore, we write $U^2$ in terms
of $Q$, and obtain precisely the $\eta L$ term in
\sprcrg.

\newsec{Discussion}

We have seen that for quantum theories that are obtained as a
decoupled Higgs branch there is a natural change of variables to the
Coulomb branch. When these theories are viewed as DLCQ descriptions of
decoupled theories in M/String theory, then this change of variables
is nothing but the AdS/CFT correspondence (and its generalizations to
non-conformal cases). This is not surprising since the Higgs branch
quantum mechanics is believed to be the full DLCQ of these theories,
but what is a pleasant surprise is that this change of variables, and
hence the AdS/CFT correspondence, is fairly straightforward in terms
of the decoupling process.

Since the full DLCQ is valid in all cases, even large curvature or
large higher genus corrections, we can use it to explore gravity in
these cases. The most striking feature that we have found in this
paper is that in the case of $AdS_7\times S^4$ the 4-sphere becomes a
non-commutative space (we have shown this for the $N_0=1$ case but we
expect it to be true more generally). It is known that open strings
generate a non-commutative structure when turning on field strengths
on D-branes\foot{One usually turns on a $B$ field in the bulk but we
can gauge it to a field strength on the brane.} \ncg, but the status
of turning on field strengths in the closed string/M-theory sector
(which in our case is a flux on $S^4$) was less clear. Our results
suggest that in this case also a similar fuzziness of space appears.

\bigbreak\bigskip\bigskip\centerline{{\bf Acknowledgments}}\nobreak

We would like to thank O. Aharony, A. Kapustin, M. Rozali and N. Seiberg 
for useful discussions. The work of MB is supported by NSF grant 
PHY-9513835, and of H.V. by NSF-grant 98-02484, a Pionier fellowship of
NWO, and the Packard foundation.

\newsec{Appendix 1.}

In this appendix we discuss a particularly simple case of reducing a
model. This will serve to explain which operations are allowed in the
reduced model and which are not.

The toy model contains 3 real bosonic coordinates and 3 complex
fermionic coordinates $X^{1,2,3},\psi^{1,2,3},{\bar\psi}^{1,2,3}$ and
is invariant under the obvious $SO(3)$:
$$Q=\psi^i\partial_i,\ {\bar Q}={\bar\psi}^i\partial_i$$
The $SO(3)$ generators are 
$$M^{ij}=X^i\partial_j-X^j\partial_i+
\psi^i{\bar\psi}^j-\psi^j{\bar\psi}^i$$
We are interested in reducing the model with respect $SO(3)$. 
The wave functions satisfy $M^{ij}\Psi=0$.
This implies that for $j=2,3$:
$$\partial_j = {1\over X^1}\bigl (X^j\partial_1-
\psi^1{\bar\psi}^j+\psi^j{\bar\psi^1}\bigr)$$
Since we will insert this expression every time there will be a
derivative $\partial_j$ we can take the limit $X^2,X^3=0$ without
worry. (Note for example that we can not use this formula to compute,
for example, $[\partial_i, X^j]$ because $X^j$ is not $SO(3)$
invariant. But we can compute things like $[\partial_j, X^2]=2X^j$).

This gives us supercharges
$$Q=\psi^1\partial_1 -\psi^j{1\over X^1}
\bigl(\psi^1{\bar\psi}^j-\psi^j{\bar\psi}^1\bigr)=
\psi^1(\partial_1+\psi^j{\bar\psi^j})$$ 
$${\bar Q}=\psi^1\partial_1 -{\bar\psi}^j{1\over X^1}
\bigl(\psi^1{\bar\psi}^j-\psi^j{\bar\psi}^1\bigr)=
{\bar\psi}^1(\partial_1-\psi^j{\bar\psi}^j)$$

It is easy to verify that $\{Q,Q\}=0$. To evaluate the Hamiltonian we
anti-commute $\{Q,{\bar Q}\}$. Even though we started with an
Hamiltonian that did not contain fermion oscillators, the new
Hamiltonian does. Still, the change is only the generation of new
3-fermion terms in the supercharges.

\newsec{Appendix 2}

In this appendix we will (partially) determine the supercharges of the
reduced Higgs branch quantum mechanics. One can compute the
supercharges of the full ADHM sigma model and then compute the reduced
supercharges. Alternatively, since the supercharges are determined by
the (super)symmetries of the problem, one can compute them in the
reduced model directly, which is the route we will take.  We will show
that the supercharges are (for some conventions see appendix 3)
\eqn\appspr{Q^\alpha_A=\eta^\alpha_A
\biggl({\partial\over\partial Q}+{a\over Q}\biggr)-
{2\over Q}\et{\al}{B}L^{BC}J_{AC}+
{1\over Q}M^{\al\be}\et{\ga}{A}\epsilon_{\be\ga},}
where $a$ is a specific number which we have not determined. 
The different operators in \appspr\ are the following:

\noindent 1. $$M^{\al\be}=J^{BB'}\et{\al}{B}\et{\be}{B'}$$ denote 
the generators of $SU(2)_R$ in the $\eta$ sector. These operators are
symmetric under the exchange $\al\leftrightarrow\be$ and satisfy
$$[M^{\al\be},M^{\al'\be'}]=
 \epsilon^{\al\al'}M^{\be\be'}+\epsilon^{\al\be'}M^{\be\al'}
+\epsilon^{\be\al'}M^{\al\be'}+\epsilon^{\be\be'}M^{\al\al'}$$
$$[M^{\al\be},\et{\ga}{A}]=\epsilon^{\al\ga}\et{\be}{A}+
			   \epsilon^{\be\ga}\et{\al}{A}$$

\noindent 2. $$L^{AB}=\nu^A_k{\bar\nu}^k_CJ^{BC}+
	   \nu^B_k{\bar\nu}^k_CJ^{AC}$$ are the generators of $USp(2)$
on the $\nu$ sector.  They satisfy
$$[L^{AB},L^{A'B'}]=
J^{AA'}L^{BB'}+J^{AB'}L^{BA'}+J^{BA'}L^{AB'}+J^{BB'}L^{AA'}.$$

As usual in the Hamiltonian formalism one first writes down the most
general supercharge and then fixes it by requiring closure
on the Hamiltonian. The most general supercharge (setting the
coefficient of derivative term to 1) is given by
$$\et{\al}{A}\biggl({\partial\over\partial Q}+{a\over Q}\biggr)+
{c_1\over Q}\et{\al}{B}\nu^B_k{\bar\nu}^k_A+ {c_2\over
Q}\eta^\alpha_B\nu^C_k{\bar\nu}^k_DJ^{BD}J_{AC} +{b\over
Q}M^{\al\be}\et{\ga}{A}\epsilon_{\be\ga}.$$ This Hamiltonian is
determined in the following way
\item{1.} The powers of $Q$ in front
of each term are determined by scale invariance under which the fermion
operators have dimension 0.
\item{2.} One might think to add a term 
${1\over Q}\et{\al}{A}\nu^D_k{\bar\nu}^k_D$, but as explained above,
this number is a constant in the relevant sector and therefore can be
absorbed into $a$.
\item{3.} Otherwise one uses $SU(2)_R\times USp(2)\times SU(N)$ 
symmetry where the last component is the remaining flavor symmetry after 
reducing the model. 

We will be interested in computing the 4-Fermi term in the
anti-commutator of two supercharges, and require that the result is a
singlet under all the global symmetries. This will determine\foot{We
have not calculated the 2-Fermi term, hence we can not compute
$a$. The system of equations for the other coefficient is
over-determined} $c_1,c_2$ and $b$. The 4-Fermi terms that we might
generate are $\eta\eta\eta\eta$, $\nu{\bar\nu}\nu{\bar\nu}$ and
$\eta\eta\nu{\bar\nu}$.

Since we will be interested in 4-Fermi terms we can anti-commute
fermion freely, neglecting the C-number that results, since it will
appear as part of the 2-Fermi term.

\medskip

\noindent{\it The $\eta\eta\eta\eta$ term}

\medskip

The relevant terms that contribute to the 4-$\eta$ term are
$$\{  \et{\al}{A}\partial_Q,
		{b\over Q}M^{\al'\be'}\et{\ga'}{A'}\eps_{\be'\ga'}  \}+
  \{  {b\over Q}M^{\al\be}\et{\ga}{A}\eps_{\be\ga},
      \et{\al'}{A'}\partial_Q					    \}+$$
$$\{ {b\over Q}M^{\al\be}\et{\ga}{A}\eps_{\be\ga}, {b\over
      Q}M^{\al'\be'}\et{\ga'}{A'}\eps_{\be'\ga'} \}\cong$$

$$\cong{-b\over Q^2} \bigl(
\et{\al}{A}M^{\al'\be'}\et{\ga'}{A'}\epsilon_{\be'\ga'}
+ ( (A,\alpha)\leftrightarrow (A',\alpha') ) \bigr)+$$
$$+{b^2\over Q^2}\biggl(
\epsilon^{\al\al'}\epsilon_{\be\ga}\epsilon_{\be'\ga'}
	M^{\be\be'}\et{\ga}{A}\et{ga'}{A'}+
M^{\al\al'}(\epsilon_{\ga\ga'}\et{\ga}{A}\et{\ga'}{A'})\biggr)$$ Where
$\cong$ denotes that we have neglected terms that are already singlets
under $SU(2)_R\times USp(2)$.

This expression is an $SU(2)_R$ singlet, in which case it is either in
the ${\bf 1}$ or ${\bf 5}$ of $USp(2)$, and an $SU(2)_R$ triplet, in
which case it is in the ${\bf 10}$ of $USp(2)$. We require the
cancelation of the $({\bf 1,5})$ and $({\bf 3,10})$ terms.

To extract the $({\bf 1,5})$ terms we contract the expression with
$\eps_{\al\al'}$ and obtain
$$-{2b\over Q^2}
\bigl(-\epsilon_{\delta\delta'}\epsilon_{\be'\ga'}
M^{\delta'\be'}\et{\delta}{A}\et{\ga'}{A'}\bigr)$$
$$-{2b^2\over Q^2} \epsilon_{\be\ga}\epsilon_{\be'\ga'}
M^{\be\be'}\et{\ga}{A}\et{\ga'}{A'}$$
For this term to cancel we require
$$b=1.$$ The solution $b=0$ is of course also a valid solution,
however, one easily sees that this can not be the right solution -
already at $N_4=2$ we have a 3-$\eta$ term in the supercharges.

To extract the $({\bf 3,10})$ term we symmetrize $A\leftrightarrow A'$
and obtain
$$-{b\over Q^2}M^{\al'\al}\eps_{\ga\ga'}\et{\ga}{A}\et{\ga'}{A'}
+{b^2\over Q^2}M^{\al\al'}\eps_{\ga\ga'}\et{\ga}{A}\et{\ga'}{A'}$$
which again gives us $b=1$.

\medskip

\noindent{\it The $\nu{\bar\nu}\nu{\bar\nu}$ term}

\medskip

The anti-commutator that contributes to the 4-$\nu$ is
$$\{ 
{c_1\over Q}\et{\al}{B}\nu^B_k{\bar\nu}^k_A+ 
{c_2\over Q}\et{\al}{B}\nu^C_k{\bar\nu}^k_DJ^{BD}J_{AC},
{c_1\over Q}\et{\al'}{B'}\nu^{B'}_{k'}{\bar\nu}^{k'}_{A'}+ 
{c_2\over Q}\et{\al'}{B'}\nu^{C'}_{k'}{\bar\nu}^{k'}_{D'}J^{B'D'}J_{A'C'}
\}$$
It is clear that the expression is symmetric to the exchange
$(\alpha,A)\leftrightarrow (\alpha',A')$ and it is also clear that the
result will be proportional to a singlet of $SU(2)_R$, i.e., to
$\epsilon^{\alpha\alpha'}$. Hence the result will antisymmetric under
$A\leftrightarrow A'$. Under $USp(2)$ the expression is either a ${\bf
1}$ or a ${\bf 5}$.

This anticommutator is
$${c_1^2\over Q^2}\biggl(
J_{BB'}\bigl(\nu^B_k{\bar\nu}^k_A\nu^{B'}_{k'}{\bar\nu}^{k'}_{A'}\bigr)
\biggr)+$$
$$+{c_1c_2\over Q^2}\biggl(
J_{A'C'}\nu^B_k{\bar\nu}^k_A   \nu^{C'}_{k'}{\bar\nu}^{k'}_B-
J_{AC'} \nu^B_k{\bar\nu}^k_{A'}\nu^{C'}_{k'}{\bar\nu}^{k'}_B
      \biggr)+$$
$$+{c_2^2\over Q^2}\biggl(
J_{AC}J_{A'C'} \nu^C_k{\bar\nu}^k_D\nu^{C'}_{k'}{\bar\nu}^{k'}_B J^{BD}
\biggr)$$

We would like to ask what are the conditions on $c_1$ and $c_2$ such
that this expression is a singlet of $USp(2)$.  To do this, its enough
to examine special cases. for example $A=1,A'=3$. In this case we
obtain the expression ($J^{12}=J^{34}=1$)
$$-{c_1^2\over Q^2}\bigl(J_{BB'}\nu^B_k\nu^{B'}_{k'}\bigr)
					{\bar\nu}^k_1{\bar\nu}^{k'}_3-
{c_1c_2\over Q^2}\bigl(\nu^B_k{\bar\nu}^{k'}_B\bigr)
	\bigl({\bar\nu}^k_1\nu^4_{k'}-{\bar\nu}^k_3\nu^2_{k'}\bigr)$$
$$+{c_2^2\over Q^2} \bigl({\bar\nu}^k_D{\bar\nu}^{k'}_BJ^{BD}\bigr)
\nu^2_k\nu^4_{k'}$$
Each term in this expression contain either $\nu^2{\bar\nu}_3$ or
$\nu^4{\bar\nu_1}$. To show under what conditions its zero its enough
to focus on terms containing one of this 2-fermion elements, say
$\nu^2{\bar\nu}^3$. The terms containing the 2-fermion element are
$$-{1\over Q^2}\bigl( 
c_1^2\nu^1_k{\bar\nu}^k_1 + c_2^2\nu^4_k{\bar\nu}^k_4 
\bigr) \nu^2_{k'}{\bar\nu}_3^{k'}-{1\over Q^2}\bigl( 
c_1^2\nu^1_{k'}{\bar\nu}^k_1 + c_2^2\nu^4_{k'}{\bar\nu}^k_4 
\bigr) \nu^2_k{\bar\nu}_3^{k'}-$$
$$-{c_1c_2\over Q^2}\bigl(\nu^B_k{\bar\nu}^{k'}_B\bigr)
{\bar\nu}^k_3\nu^2_{k'}$$

There are several ways for this to be zero. The simple one is to set
$c_1+c_2=0$. In this case the expression vanishes. This is however not
the only way. Another way is to set $c_1=c_2$.  The reason is that we
do not really need to require that the expression vanishes
identically, rather it should vanish only on the flavor invariant
gauge invariant states in our Hilbert space. As explained in section
{\it 4.2} these states are of the form
\eqn\sttagn{\epsilon^{l_1..l_{N}}\epsilon^{t_1..t_{N}}
\nu^{A_1}_{l_1}....\nu^{A_{N}}_{l_{N}}
\nu^{B_1}_{t_1}....\nu^{B_{N}}_{t_{N}}\wvfnc{\phi}} 
On these states $\nu^B_k\nu^{k'}_B\propto \delta^{k'}_k$ (with a fixed
coefficient). Therefore terms containing $\nu^B_k\nu^{k'}_B$ are not
really 4-Fermi terms. Hence the term which is proportional to $c_1c_2$
does not contribute. Similarly we can substitute (setting $c_1=c_2$)
$$ \nu^1_{k'}{\bar\nu}^k_1+\nu^4_{k'}{\bar\nu}^k_4=
  -\nu^2_{k'}{\bar\nu}^k_2+\nu^3_{k'}{\bar\nu}^k_3\ +\ fixed\ number$$
to obtain that the total expression is
$$-{c^2\over Q^2}
(\nu^B_{k'}{\bar\nu}^k_B)\nu^2_{k'}{\bar\nu}^{k'}_3$$ which is again a
2-Fermi operator on the relevant Hilbert space. 

The solution which is correct is actually $c_1=c_2$. In this case, the
two $\eta\nu{\bar\nu}$ terms combine to give
$$c\et{\al}{B}L^{BC}J_{AC}$$

Be before we proceed it is worth returning to the $USp(2)\times
SU(2)_R$ singlet terms which we have been neglecting left and
right. These terms do not give us any restrictions on the supercharges
but they are part of the Hamiltonian. The $\nu^2{\bar\nu}^2$ term in
the Hamiltonian turns out to be proportional to $$J^{AA'}J_{BB'}
\bigl(\nu^B_k{\bar\nu}^k_A\nu^{B'}_{k'}{\bar\nu}^{k'}_{A'}\bigr)$$, 
which is in turn proportional to the $USp(2)$ 2nd Casimir (up to 2
Fermi terms which we have not calculated).

\medskip

\noindent{The $\eta\eta\nu{\bar\nu}$ term}

\medskip

The anti-commutators which can contribute to such a term are:

\noindent I. $\{ \et{\al}{A}\partial_Q, 
                 {c\over Q}\et{\al'}{B'}{L(\nu)}^{B'C'}J_{C'A'} \} + 
(A,\alpha)\leftrightarrow (A',\alpha') $

\noindent II. $\{ {c\over Q}\et{\al}{B}{L(\nu)}^{BC}J_{AC},
  	           {c\over Q}\et{\al'}{B'}{L(\nu)}^{B'C'}J_{A'C'} \}$

\noindent III. $\{ {c\over Q}\et{\al}{B}{L(\nu)}^{BC}J_{AC},
	  {1\over Q}M^{\al'\be'}\et{\ga'}{A'}\epsilon_{\be'\ga'} \}+ 
(A,\alpha)\leftrightarrow (A',\alpha') $

Evaluating these terms we obtain

\noindent I. ${-c\over Q^2} ( \et{\al}{A}\et{\al'}{B}L^{BC}J_{A'C}+
		  \et{\al'}{A'}\et{\al}{B}L^{BC}J_{AC} )$


\noindent II.
${c^2\over Q^2}\biggl(
(-J^{B'B}\et{\al}{B}\et{\al'}{B'}L^{CC'}J_{AC}J_{A'C'}+
\et{\al}{B}\et{\al'}{B'}L^{BB'}J_{A'A}$
$$-\et{\al}{A'}\et{\al'}{B'}L^{CB'}J_{AC}
  +\et{\al}{B}\et{\al'}{A}L^{BC'}J_{A'C'}\biggr)$$

\noindent III.
${bc\over Q^2} \biggl(\eps^{\al\al'}\et{\be'}{B}\et{\ga'}{A'}
				\eps_{\be'\ga'}L^{BC}J_{AC}
				+\et{\al'}{B}\et{\al}{A'}L^{BC}J_{AC}-
				M^{\al'\al}J_{BA'}L^{BC}J_{AC}\biggr)+
				(\alpha,A)\leftrightarrow (\alpha',A')
				$


The ${\bf (1,5)}$ under $SU(2)\times USp(2)$ part is:
$${1\over Q^2}(2c+c^2)
\bigl( \et{\al}{A} \eps_{\al\al'} \et{\al'}{B} L^{BC} J_{A'C} -
       \et{\al}{A'}\eps_{\al\al'} \et{\al'}{B} L^{BC} J_{AC}\bigr)$$
giving us $c=-2$.

To examine the ${\bf (3,10)}$ part we write the expression for
$\al=\al'=0$, we then get:
$${1\over Q^2}(-2c-c^2)
\bigl(
\et{0}{A}\et{0}{B}L^{BC}J_{A'C}+\et{0}{A'}\et{0}{B}L^{BC}J_{AC}
\bigr)$$
giving us again $c=-2$.
 
\listrefs 

\end